\documentclass[
	a4paper, % Paper size, use either a4paper or letterpaper
	10pt, % Default font size, can also use 11pt or 12pt, although this is not recommended
	unnumberedsections, % Comment to enable section numbering
	twoside, % Two side traditional mode where headers and footers change between odd and even pages, comment this option to make them fixed
]{LTJournalArticle}

\usepackage{svg}
\runninghead{} % A shortened article title to appear in the running head, leave this command empty for no running head

\footertext{\textit{RISC-V Summit Europe, Paris, 12-15th May 2025}} % Text to appear in the footer, leave this command empty for no footer text

\title{REPTILES: Repeated Tiles of Sargantana,\\a RISC-V multicore based on OpenPiton} % Article title, use manual lines breaks (\\) to beautify the layout

\author{%
Noelia Oliete-Escuín\textsuperscript{1}%
\thanks{Corresponding author: Barcelona Supercomputing Center, e-mail: {\tt noelia.oliete@bsc.es}},
Arnau Bigas-Soldevila\textsuperscript{1}, 
Narcís Rodas\textsuperscript{1}, 
Albert Aguilera\textsuperscript{1}, 
Sajjad Ahmad\textsuperscript{1}, \and
Jonathan Balkind\textsuperscript{2}
\thanks{University of California, Santa Barbara},
Xavier Carril\textsuperscript{1},
Max Doblas\textsuperscript{1}, 
Ivan Díaz\textsuperscript{1}, 
Roger Figueras\textsuperscript{1}, 
Alireza Foroodnia\textsuperscript{1}, \and 
Cesar Fuget\textsuperscript{3}
\thanks{University Grenoble Alpes, Inria},
Ignacio Genovese\textsuperscript{1}, 
Raúl Gilabert\textsuperscript{1}, 
Abbas Haghi\textsuperscript{1}, 
Alexander Kropotov\textsuperscript{1}, 
Neiel Leyva\textsuperscript{1}, \and
Oscar Lostes-Cazorla\textsuperscript{1}, 
Lorién López-Villellas\textsuperscript{4}
\thanks{University Zaragoza},
Davy Million\textsuperscript{5}
\thanks{University Grenoble Alpes, CEA},
Alireza Monemi\textsuperscript{1}, 
Sérik Pérez\textsuperscript{1}, \and
Juan Antonio Rodríguez\textsuperscript{1},  
Víctor Soria-Pardos\textsuperscript{1}, 
Behzad Salami\textsuperscript{1}, 
Francesc Moll\textsuperscript{1}, 
Oscar Palomar\textsuperscript{1}, \and
Miquel Moretó\textsuperscript{1} and Lluc Alvarez\textsuperscript{1}
}
\date{12-15th May 2025}
\renewcommand{\maketitlehookd}{%
	\begin{abstract}
		\noindent Chip industry continues advancing and expanding modern computing systems, resulting in more complex multi-core processors. Conversely, academic projects face scalability challenges due to limited resources, highlighting the need for open-source frameworks that enable innovation and knowledge sharing. Recently, several open-source proposals have emerged, offering flexible and scalable designs, but fail to meet the performance demands of modern High-Performance Computing (HPC) applications. In this project, we present REPTILES, an open-source RISC-V multicore framework based on OpenPiton\thanks{REPTILES open-source code: \url{https://github.com/bsc-loca/openpiton}}. REPTILES interconnects multiple Sargantana cores with the memory hierarchy of OpenPiton. Moreover, we present the new features incorporated in Sargantana and OpenPiton designs to improve the performance of HPC applications. We demonstrate that REPTILES presents suitable scalability, achieving a speedup of 3.1$\times$ on average with 4 cores. Additionally, we show that Sargantana's new features increase the performance of vector addition benchmark in a 9.3$\times$.   
	\end{abstract}
}

\begin{document}

\maketitle % Output the title section

%----------------------------------------------------------------------------------------
%	ARTICLE CONTENTS
%----------------------------------------------------------------------------------------

%Paper structure:
%\begin{enumerate}
%    \item Introduction: 0.25 pages
%    \item System Overview and Features: 0.5 pages
%    \item OpenPiton Improvements and New Open-Sourced Features: 0.5 pages
%    \item Sargantana Improvements and New Open-Sourced Features: 0.25 pages
%    \item System Evaluation: 0.5 pages
% \item Conclusions + Biblio: 0.25 pages
%  \item \textbf{TOTAL}: 0.25+0.5+0.5+0.25+0.5+0.25=2.25 pages...
%\end{enumerate}
\vspace{-0.6cm}
\section{Introduction}
\vspace{-0.3cm}
% Main Idea 
% We need open-source
The chip industry continues developing and scaling modern systems, leading to increasingly complex multi-core processors. However, academic projects struggle to achieve similar scalability due to limited resources and infrastructure. To overcome these challenges, the community needs open-source architecture frameworks where researchers can explore and design their ideas. Open-source frameworks provide flexibility and transparency, fostering innovation and collaboration. 

% Problem We need HPC open-source
In recent years, different proposals have emerged to provide open-source solutions. Nevertheless, with the growing demand for High-Performance Computing (HPC), current open-source projects struggle to meet the performance requirements of modern applications. 

% Our proposal
In this project, we present \textbf{REPTILES}, \textbf{REP}eated \textbf{TILE}s of \textbf{S}argantana, a RISC-V multicore based on OpenPiton. REPTILES is an open-source multi-core architecture that aims to provide an accessible HPC framework where researchers can develop, experiment with, and optimize HPC applications. Additionally, we detail the high-performance features included in the design.  

% Key results
We show that REPTILES presents strong scalability, achieving an average speedup of 3.1$\times$ with 4 cores. Furthermore, we demonstrate that Sargantana's new features enhance the performance of the vector addition benchmark by 9.3$\times$.

%------------------------------------------------
\vspace{-0.7cm}
\section{System Overview and Features}
\vspace{-0.3cm}
%- Overview multicore
% - 2, 4 cores en FPGA
%  - Mides caches, etc
%  - Ethernet
The multicore system presented in this poster is based on OpenPiton and Sargantana, publicly available open-source designs based on Verilog and SystemVerilog.
OpenPiton is a multi-core framework. 
%that offers a highly configurable memory hierarchy.
%supporting various cache sizes, associativity levels, core counts, and other tunable parameters. 
The cores themselves are replicated Sargantana tiles, single-issue in-order RISC-V 64-bit processors.
%The system we present integrates and improves these two open-source projects.
%We evaluate REPTILES in RTL simulation and FPGA. The simulated system includes 4 Sargantana cores with 128-bit RVV 1.0 SIMD units and a cache hierarchy composed of an in-house 32~KB L1 instruction cache, a 32~KB High-Performance L1 data cache, a 128~KB L1.5 cache, and a 256~KB shared L2 cache. Additionally, the private cache levels are configured with 64 MSHRs and connected via 512-bit NoC buses to the L2 cache. The FPGA prototype also includes 4 Sargantana cores. However, due to resource limitations of the Alveo u55c FPGA, it excludes the SIMD unit, and both the cache sizes and the NoC width are reduced. As a result, the private L1 and L1.5 caches are configured as 8~KB caches, the shared L2 cache as 64~KB, and the NoC width is reduced to 64~bits.

We evaluate REPTILES in an FPGA prototype that includes 4 Sargantana cores, without SIMD unit, and a cache hierarchy composed of an in-house 16~KB L1 instruction cache, a 16~KB High-Performance L1 data cache, a 32~KB L1.5 cache, and a 64~KB shared L2 cache with 64~B cache blocks. The private cache levels are configured with 64 MSHRs and connected via 64-bit NoC buses to the L2 cache. The evaluated configuration is limited due to the resource limitations of the Alveo u55 FPGA.
In addition, the FPGA prototype includes 16~GB of HBM main memory and Ethernet support. These features enable the use of a Software Development Vehicle (SDV) where a shared filesystem can be mounted, facilitating extensive benchmarking and live interactive demos.

%------------------------------------------------
\vspace{-0.6cm}
\section{OpenPiton Improvements and New Open-Sourced Features}
%- Sistema basat en OpenPiton
  %- HPDC Integration (Sargantana/Ariane?)
  %- NoCs 64 bits -> 512
  %- Cachelines 16B -> 64B 
  %- MSHRs 
  %- Multiple Memory Controllers??? @LluC
% Paragraph 1: OpenPiton motivation open sourced + performanc limitations
\vspace{-0.3cm}
The increasing need for HPC calls for open-source solutions. Initiatives like OpenPiton~\cite{10.1145/2980024.2872414} have been developed to support scalable and customizable architectures. Nevertheless, these frameworks often face performance limitations that restrict their capability to handle compute-intensive workloads effectively. 

% Paragraph 2: OpenPiton introduction
%OpenPiton was initially designed to support SPARC v9 architectures (OpenSPARC T1). Lately, the platform has been modified to operate in up to four different architectures, such as x86 and RISC-V architectures (RISC-V 32-bit, and RISC-V 64-bit), enabling a heterogeneous system. 
OpenPiton operates in different architectures, such as SPARC v9, x86 and RISC-V architectures (RISC-V 32-bit and RISC-V 64-bit).
% Paragraph 3: OpenPiton baseline architecture
The architecture of OpenPiton consists of a chipset and one or more tiles. The chipset includes modules that connect the tiles with the peripherals, such as the UART. Each tile integrates three NoC routers, the core, and the cache hierarchy. This hierarchy includes private L1 instruction and data caches, a private L1.5 cache, and a shared distributed L2 cache that implements a directory-based MESI coherence protocol. Although OpenPiton provides numerous advantages, it encounters performance constraints that limit its suitability for HPC domains.

% Paragraph 4: New features
% TODO: Añadir todas las features (L2 mshrs, associativity, noc width 704b single-flit transactions, parallel-srams)
%Recent work proposes a set of improvements to the NoC and the memory hierarchy to upgrade OpenPiton to meet the HPC requirements~\cite{10599264}. In this work, we integrate these features in our design, specifically, the NoC width increase from 64 bits to 512 bits, the support for configurable cache block sizes (64, 32, and 16 bytes) for all the cache levels, and parametric number of MSHRs for the L1.5 cache. Additionally, we enhance the Sargantana core by including a High-Performance Data Cache (HPDcache)~\cite{10.1145/3587135.3591413} in our design. 

Recent work proposes a set of improvements to the NoC and the memory hierarchy to upgrade OpenPiton to meet the HPC requirements~\cite{10599264}. In this work, we integrate these features into our design and introduce new ones. Specifically, the parametric NoC width from 64 bits up to 704 bits, the support for configurable cache block sizes (64, 32, and 16 bytes) for all the cache levels and the parametric number of MSHRs, associativity and the parallel SRAM access for the L1.5 and L2 caches. Additionally, we enhance our design with the connection of Sargantana with the High-Performance Data Cache (HPDcache)~\cite{10.1145/3587135.3591413}.

%------------------------------------------------
\vspace{-0.6cm}
\section{Sargantana Improvements and New Open-Sourced Features}
\vspace{-0.3cm}
%- Millores Sargantana
%  - Noves instruccions SIMD
%  - Bugfixes pel tapeout 
Sargantana is a Linux-capable 64-bit RISC-V processor that implements the RV64G ISA and achieves a 1.26 GHz frequency using a 22nm technology node \cite{sargantana2022}. Since its open-source release, it has received several improvements, such as architecture support for more RISC-V extensions and general usability enhancements to the RTL simulation environment.

The most significant change has been the upgrade from the RISC-V Vector Extension (RVV) 0.7 version to 1.0. In \cite{sargantana2022}, Sargantana only supported a small set of arithmetic vector instructions that could be used when manually vectorizing specific codes. Currently, the core supports most of the extension specifications, except for the LMUL>1 setting and vector FP instructions. It also implements renaming for vector configuration instructions (previously, they stalled the pipeline), leading to more remarkable performance in vectorized codes.

Other notable added extensions are Sdext for debugging support and Sscofpmf to enable reading the core performance counters in Linux via perf. 

Regarding the RTL simulation environment, we added support for the saving and restoring feature using Verilator. This allows users to periodically create checkpoints during simulation that store the model of the design in an intermediate state. Later, the simulation can be resumed from that point without re-running, which can be very helpful for debugging.

%------------------------------------------------
\vspace{-0.5cm}
\section{System Evaluation}
\vspace{-0.3cm}
%- Avaluació
%  - Openpiton cacheline 16B vs 64B, noc width 64b vs 512b
%  - Sargantana w/o SIMD
%  - Resultats NAS
%  - Resultats SPEC (dona temps?) 
  
\begin{figure}[h]
    \centering
    \includegraphics[width=\columnwidth]{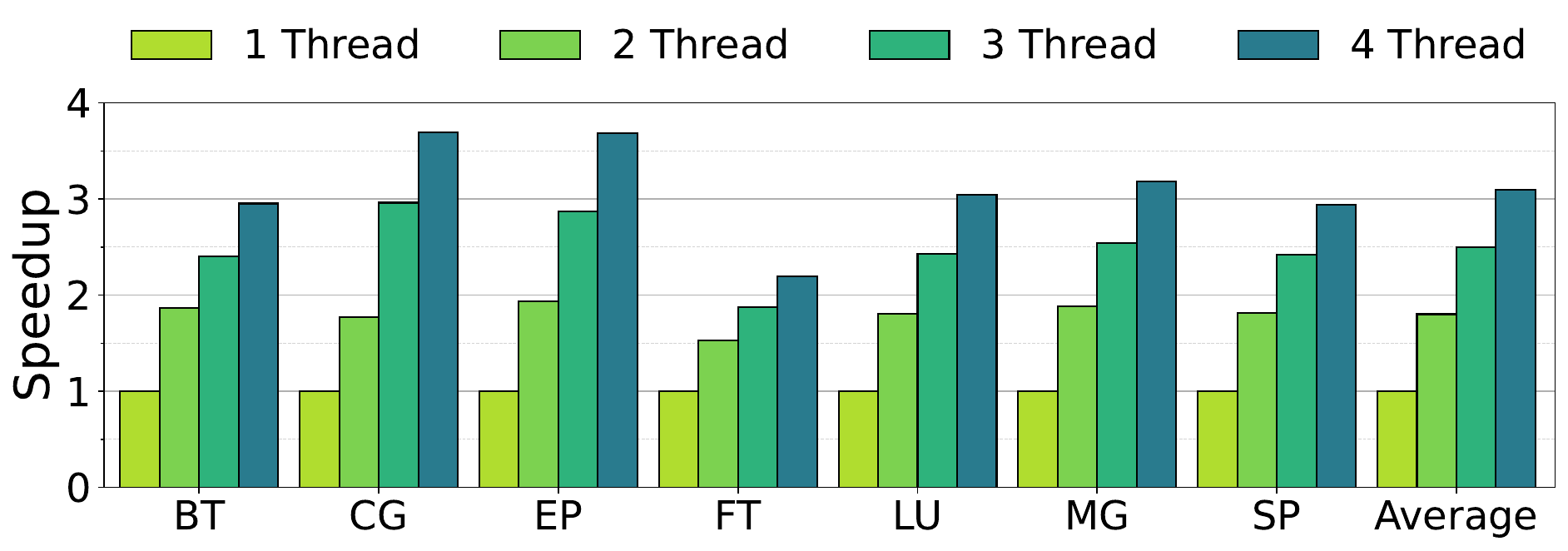}
    \caption{Execution speedup of NAS benchmarks over 2, 3, and 4 threads with respect to 1 thread.}
    \label{fig:FPGA_NAS_results}
\end{figure}

%------------------------------------------------
% TODO: Quitar simulación y mirar si faltan datos en el de FPGA. 
We evaluate the performance of REPTILES running the NAS Parallel Benchmarks with OpenMP over Linux in the FPGA system. Figure~\ref{fig:FPGA_NAS_results} shows the speedup achieved on each benchmark with 2, 3, and 4 threads with respect to 1 thread.
When increasing the number of threads, we can observe suitable scalability for all the benchmarks and significant performance improvements. The configuration with 4 threads also achieves a speedup of 3.6$\times$ for the CG and EP benchmarks and an average of 3.1$\times$. 

Additionally, we analyze the performance of Sargantana with RVV extension executing a vector addition benchmark of 8-bit elements on a standalone basis. We observe that the Sargantana RVV version achieves a 9.3$\times$ speedup with respect to the scalar version. 

\vspace{-0.3cm}
  
%----------------------------------------------------------------------------------------
%	 REFERENCES
%----------------------------------------------------------------------------------------
\vspace{-0.6cm}
\printbibliography % Output the bibliography
%----------------------------------------------------------------------------------------
\end{document}